\documentclass[seceq]{ptptex}
\newcommand{\fslash}[1]{\ooalign{\hfil/\hfil\crcr$#1$}}
\title{%        %You can use \\ for explicit line-break
Proper Construction of the Continuum in Light-cone
QCD Sum Rules
}

\author{%       %Use \scshape  for the family name
Hungchong \textsc{Kim},$^{1,2}$ Su Houng \textsc{Lee}$^{1}$
 and Makoto \textsc{Oka}$^{2}$
}

\inst{%         %Affiliation, neglected when [addenda] or [errata]
$^1$ Institute of Physics and Applied Physics, Yonsei
University,
        Seoul 120-749, Korea\\
        $^2$ Department of Physics,Tokyo
Institute of Technology, Tokyo 152-8552, Japan
}

\abst{%         %this abstract is neglected when [addenda] or [errata]
A proper method of subtracting the continuum contributions in
light-cone QCD sum rules (LCQSR) is demonstrated. Specifically, we
calculate the continuum corresponding to a typical operator product 
expansion (OPE) appearing
in LCQSR by properly combining the double dispersion relation with
QCD duality. We demonstrate how the subtraction terms can
spuriously contribute to the sum rules. In the limit of zero
external momentum, removal of the spurious continuum is found to
yield the sum rules using the single-variable dispersion relation.
The continuum factor constructed in this way differs from that 
appearing in usual LCQSR. The difference substantially affects the
extraction of hadronic parameters from the correlation function
involving baryon currents.
}

\begin{document}

\maketitle

\section{Introduction}
\label{sec:intro}

The QCD sum rule~\cite{Shifman:bx,Shifman:by} is a framework 
widely used to investigate hadronic properties in terms of QCD
degrees of freedom~\cite{Reinders:1984sr}. In this method, it is
crucial to represent a correlation function through a dispersion
relation. This is because QCD calculations of the correlation
function through the operator product expansion (OPE) can be done
only in deep space-like regions, whereas the hadronic parameters are
defined by the nonanalytic structure existing in time-like regions.
Through a dispersion relation, the calculated correlation function
can be matched with the hadronic parameters.

Within the QCD sum rule framework, a correlation function with an
external field is often considered for the purpose of calculating 
meson-baryon
couplings~\cite{Kim:1998wp,Kim:1999ue,Kim:1998ir,Kim:1999tc,Kim:2000ty,Doi:2000dx,Aliev:1999ce,Zhu:1998am},
$g_{D^* D \pi}$ and $g_{B^* B\pi}$~\cite{Belyaev:1995zk}, and
magnetic moments of baryons~\cite{Ioffe:ju,Aliev:2000rc}. At
present, there are two methods to construct QCD sum rules when
an external field is present:  the conventional approach, relying
on the short-distance expansion, and the light-cone QCD sum rule
(LCQSR),~\cite{Balitsky:ry,Braun:1988qv,Braun:1989iv,Halperin:1992gt}
based on the expansion along the light-cone. Within the
conventional approach one expands over the small momentum of the
external field and constructs a separate sum rule for the
correlation function appearing at each order of the expansion. In
this approach, however, there has been a debate over which
dispersion relation to
use;~\cite{Kim:1999rg,Ioffe:1995jt,Krippa:1999de} one can start
either from the single-variable dispersion relation or from the
double-variable dispersion relation, and the results seem to
depend strongly on which dispersion relation is
used.~\cite{Kim:1999rg}. Later, it was shown~\cite{Kim:1999ur}
that the two dispersion relations in the conventional sum rule
yield identical results provided that the spurious continuum
appearing in the double dispersion relation is properly
eliminated. The spurious continuum, which comes from subtraction
terms, appears when QCD duality is naively imposed on the double
dispersion relation.  A safer approach, therefore, is to start from the
single dispersion relation.

The spurious continuum can cause more serious problems in LCQSR, in
which the double-variable dispersion relation is the only choice
in representing correlation functions. In LCQSR, one keeps the
external momentum finite, and the correlation function contains two
momenta. The double-variable dispersion relation and, subsequently,
the double Borel transformations must be applied for the two
finite momenta. Then, it is not entirely clear how to apply QCD duality 
in subtracting the
continuum, similar terms that are
mathematically spurious can appear.

Indeed, in the case of the $D^* D \pi$ coupling, the issue of the
spurious continuum has been briefly investigated~\cite{Kim:2002xc} in
an attempt to revolve the discrepancy between the existing LCQSR
calculation~\cite{Belyaev:1995zk} and the recent CLEO
measurement~\cite{Anastassov:2001cw}. In particular, the
LCQSR~\cite{Belyaev:1995zk} predicts this coupling to be around 12.5,
which is substantially lower than a recent
measurement~\cite{Anastassov:2001cw}, 17.9. The experimental value,
however, is consistent with the quark model,~\cite{Becirevic:1999fr}
and lattice calculation~\cite{Abada:2002xe}.  This clearly
indicates that the existing LCQSR need to be improved. It is,
however, noteworthy that
the coupling calculated from the LCQSR with the continuum
modification~\cite{Kim:2002xc} are comparable
to the experimental value. In this work, we now generalize the
argument of Ref.~\citen{Kim:2002xc} and propose an effective method
of subtracting the continuum in the LCQSR. This should provide a
guideline for future applications of LCQSR. We study how this
improvement changes the existing results of LCQSR.

\section{A proper method of subtracting the continuum in LCQSR}
\label{sec:duality}

Most LCQSR consider two-point correlation functions with an
external field. A typical OPE that often appears in LCQSR takes
the form~\cite{Zhu:1998am}
\begin{eqnarray}
\int^1_0 dv~\varphi (v) \Gamma \left ( {D\over 2} - n \right )
\left [ {1\over -vp_1^2 - (1-v)p_2^2} \right ]^{{D\over 2}-n}\ .
\label{1st}
\end{eqnarray}
Here $D$ is the dimensionality of the space. The wave function of the
external field is denoted by $\varphi (v)$. When the external
field is a pion, $\varphi (v)$ is the pion wave function,
while its argument $v$ represents the fractional momentum carried
by a quark inside the pion. The wave function is usually a
polynomial in $v$ (for the case of a pion external field, see
Ref.~\citen{Belyaev:1995zk}.):
\begin{eqnarray}
\varphi (v) = \sum_{k} a_k~v^k\ .
\label{poly}
\end{eqnarray}
Thus, it would be sufficient to consider the case $\varphi(v)
\rightarrow v^k$. For notational simplicity, let us introduce
\begin{eqnarray}
A\equiv vp_1^2 + (1-v)p_2^2\ .
\end{eqnarray}
Since we are concerned with the duality issue in subtracting the
continuum in LCQSR, we focus on the OPE that is dual to the
continuum. Because the dimensionality of space satisfies $D \sim 4$, 
the typical
OPE, Eq.~(\ref{1st}), has a pole at the zeros of $A$ for $n \le 1$.
Therefore the OPE with $n \le 1$ should not be dual to the
continuum.~\footnote{Note, however, that in the case $n=1$,
the continuum contribution is accounted for in normal LCQSR but,for 
$D^*D\pi$ coupling, it has been discussed~~\cite{Kim:2002xc} that
the case $n=1$ should not be dual to the continuum. Indeed,
the LCQSR without the continuum
in this case provide a coupling comparable to the experiment value.}

For $n\ge 2$, $\Gamma \left ( {D\over 2} - n \right )$ is
singular. In this case, one expands around $2-D/2 \equiv
\epsilon\rightarrow 0$~\cite{Novikov:gd} to separate the regular
part from the singular part,
\begin{eqnarray}
\Gamma \left ( {D\over 2} - n \right )
\left [ 1\over -A \right ]^{{D\over 2}-n}
\sim
-A^{n-2} {\rm ln}(-A) {1 \over (n-2)~!}
+ { \Gamma (\epsilon) \over (n-2)~!} A^{n-2} \ ,
\label{ttt}
\end{eqnarray}
so that our trial OPE (with $\varphi (v) \rightarrow v^k$) becomes
\begin{eqnarray}
\Pi^{\rm ope}(p_1^2,p_2^2) \equiv &-& \int^1_0 dv~  {v^k \over (n-2)~!}
~A^{n-2}
~{\rm ln} (-A)
\nonumber \\
&+& \int^1_0 dv~  {v^k
\over (n-2)~!} \Gamma (\epsilon) A^{n-2} \ .
%\nonumber \\
%&&
~~(n\ge 2, k\ge 0)\ 
\label{tope}
\end{eqnarray}
We note here that the second term contains the singular
coefficient $\Gamma (\epsilon)$. However, it is a simple power of
$A$, constituting the so called ``subtraction terms'' in QCD sum rules.
To get the finite result, it is necessary to eliminate the second
term.  This is one important reason to apply the Borel
transformations, ${\cal B} (M_1^2,-p_1^2) {\cal B}
(M_2^2,-p_2^2)$, where
\begin{eqnarray}
{\cal B} (M^2, Q^2) = \lim_{Q^2,n\rightarrow \infty, Q^2/n=M^2}
{(Q^2)^{n+1}\over n!} \left ( -{d\over dQ^2} \right )^n\ .
\end{eqnarray}
Under this operation, the subtraction terms containing $\Gamma
(\epsilon)$ vanish.

For the OPE given by Eq.~(\ref{tope}), let us illustrate how a sum
rule is constructed on the OPE side. The first
step is to obtain the spectral density $\rho^{\rm ope} (s_1, s_2)$
corresponding to the OPE (the logarithmic part) through the double 
dispersion relation
\begin{eqnarray}
\Pi^{\rm ope} (p^2_1, p^2_2) = 
\int^\infty_0 d s_1 \int^\infty_0
ds_2 { \rho^{\rm ope} (s_1, s_2) \over (s_1 - p^2_1) (s_2 -
p^2_2)}\ . \label{double}
\end{eqnarray}
Then, following the duality argument, one subtracts the continuum
contribution lying above the threshold $S_0$ simply by restricting
the integral within the range $[0, S_0]$,
\begin{eqnarray}
\Pi^{\rm ope}_{\rm sub} \equiv
\int^{S_0}_0 ds_1 \int^{S_0}_0 ds_2 {\rho^{\rm ope} (s_1, s_2)\over
(s_1-p_1^2)(s_2-p_2^2) } \ .
\label{rcont}
\end{eqnarray}
As we discuss below, however, this step is dangerous because,
there might be spurious terms contributing to the sum rule.
Nevertheless, the subsequent Borel transformations yield the
final expression for the OPE side of the sum rule,
\begin{eqnarray}
{\cal B} (M_2^2, -p_2^2){\cal B} (M_1^2,-p_1^2) \Pi^{\rm ope}_{\rm
sub} = \int^{S_0}_0 ds_1 \int^{S_0}_0 ds_2~ e^{-s_1/M_1^2
-s_2/M_2^2} \rho^{\rm ope} (s_1,s_2)\ . \label{bor}
\end{eqnarray}
We show how the spurious terms enter in this prescription
based on a simple mathematical reasoning.

To proceed, let us first demonstrate how the spectral density is normally
determined in LCQSR. A common method is to apply
the following Borel transformations to 
Eq.~(\ref{double})~\cite{Beilin:1984pf,Ball:1993xv,Nesterenko:1982gc};
\begin{eqnarray}
&&{\cal B} \left ( \tau_2^2, {1\over M_2^2} \right ){\cal B}
\left ( \tau_1^2, {1\over M_1^2} \right )
{\cal B} (M_2^2, -p_2^2){\cal B} (M_1^2,-p_1^2) \Pi^{\rm ope} (p^2_1, p^2_2)
\nonumber \\
&&=\rho^{\rm ope} \left ( {1\over \tau_1^2}, {1 \over \tau_2^2} \right )\ .
\label{spec}
\end{eqnarray}
For the OPE given in Eq.~(\ref{tope}), this equation yields
a spectral density of the
form~\cite{Belyaev:1995zk,Aliev:2000rc,Halperin:1992gt}
\begin{eqnarray}
\rho^{\rm ope} (s_1, s_2) = {s_2^{n+k-1}  \over (n+k-1)!} \left
(-{\partial \over \partial s_2} \right )^k \delta(s_1 -s_2),~{\rm
where}~ n\ge 2~,~ k \ge 0 \ . \label{fspec}
\end{eqnarray}
Note that the term containing the powers of $A$ in Eq.~(\ref{tope})
vanishes under the Borel transformations. A crucial point that we
want to make is that this spectral density satisfies the double
dispersion relation Eq.~(\ref{double}), up to {\it subtraction
terms}.  Of course, the subtraction terms do not contribute if the 
sum rule is constructed under the double dispersion relation with
the integral taken from $0$ to $\infty$. However, we emphasize that, as
this spectral density enters in the restricted interval of $[0, S_0]$, 
the subtraction terms can contribute to the sum rule. We
refer to such contributions from the subtraction terms as
``spurious''.

To demonstrate in detail, let us see how the spectral density
Eq.~(\ref{fspec}) reproduces the logarithmic part of the OPE in
Eq.~(\ref{tope}) through the double dispersion relation
Eq.~(\ref{double}). This is a natural mathematical check. We first
substitute Eq.~(\ref{fspec}) into Eq.~(\ref{double}) to obtain
\begin{eqnarray}
&&\int^\infty_0 ds_1 \int^\infty_0 ds_2 { \rho^{\rm ope} (s_1,s_2)
\over (s_1 -p_1^2)(s_2-p_2^2) } \nonumber \\
&&={1 \over (n+k-1)~!}~ \int^\infty_0 ds_1 \int^\infty_0 ds_2 {
s_2^{n+k-1} \over (s_1 - p_1^2) (s_2 -p_2^2)} \left (-{\partial
\over \partial s_2} \right )^k \delta(s_1 -s_2) \ . \label{int1}
\end{eqnarray}
We integrate first over $s_1$ by moving the part $\int^\infty_0
ds_1 /(s_1-p_1^2)$ inside the partial derivatives. After
performing the partial derivatives, the Feynman parametrization
leads to
\begin{eqnarray}
\int^1_0 dv~ v^k~ {(k+1)! \over (n+k-1)!}
 \int^\infty_0 ds_2 {s_2^{n+k-1} \over (s_2-A)^{k+2} }\ .
\label{int3}
\end{eqnarray}
Here again, we have introduced $A=vp_1^2 +(1-v)p_2^2$ for
notational simplicity. We then perform the $s_2$ integration by
parts successively and thereby reduce the power of the denominator order
by order. In this process, we eventually end up with
\begin{eqnarray}
\int^1_0 dv~ v^k~ \left [ \int^\infty_0 ds_2 {1 \over (n-2)!}
{s_2^{n-2} \over s_2-A} - \sum_{i=0}^{k} { (k-i)! \over (n+k-1-i)!
} {s_2^{n+k-1-i} \over (s_2-A)^{k+1-i} } \Bigg |^\infty_0 \right
]\ . \label{check}
\end{eqnarray}
One can see that the first term involving the $s_2$ integral is
sufficient for reproducing the logarithmic part of
Eq.~(\ref{tope}). This can be easily seen by rewriting the
numerator $s_2^{n-2} = (s_2 -A +A)^{n-2}$ and making use of the
binomial formula; that is, the double dispersion relation
Eq.~(\ref{double}) is satisfied with {\it this first term only}.
{\it We do not need the second term involving the summation to 
reproduce the OPE with which we started.}

Then, what is the nature of the second term involving the summation? 
We note that the power in the numerator is greater than or equal to 1 
because  $n \ge 2$, $k\ge 0 $ $0\le i \le k $. At the
lower limit, $s_2=0$, therefore, the second term is always 0. But
at the upper limit, $s_2= \infty$, it consists of powers of $A$ whose
coefficients are infinite. This term at the upper limit has 
precisely the same form as the powers of $A$ in
Eq.~(\ref{tope}) constituting the so-called ``subtraction terms'', and
we know that this should not contribute to the sum rule. What is
crucial here is that, when the double dispersion relation is
naively restricted within the integration interval $[0, S_0]$, the
upper limit becomes $s_2=S_0$, and the second term at this new
upper limit no longer consists of powers of $A$. Instead, it has a pole at
$A=S_0$, which obviously does not vanish even after the Borel
transformations with respect to $-p^2_1$ and $-p^2_2$. Therefore,
this contribution to the sum rule is spurious as in the case of the
subtraction terms.

Another indication that the second term of
Eq.~(\ref{check}) is spurious can be seen in the limit that the
external momentum vanishes. In this limit, we have $p_1^2 = p_2^2
\equiv q^2$, $A = q^2$ and the OPE of interest,
Eq.~(\ref{tope}), becomes
\begin{eqnarray}
\Pi^{\rm ope} (q^2)=-{1\over (k+1)(n-2)!} (q^2)^{n-2} {\rm ln}
(-q^2) + {\rm [subtraction~ terms]} \ .
\end{eqnarray}
As $\Pi^{\rm ope}(q^2)$ contains only one variable, one must use
the single-variable dispersion relation
\begin{eqnarray}
\Pi^{\rm ope} (q^2) = \int^\infty_0 ds {\rho^{\rm ope} (s) \over
s-q^2}
\end{eqnarray}
in order to obtain the corresponding spectral density. Within this
approach, we immediately find that the spectral density is given
by
\begin{eqnarray}
\rho^{\rm ope}(s)=  {1\over (k+1)(n-2)!}s^{n-2} \ ,
\end{eqnarray}
which yields the OPE side after subtracting the continuum
\begin{eqnarray}
\int^{S_0}_0 ds {1\over (k+1)(n-2)!} {s^{n-2} \over s-q^2}\ .
\end{eqnarray}
This expression should be recovered from Eq.~(\ref{check}) when we
set the external momentum to 0. In fact, the first term in
Eq.~(\ref{check}) can reproduce this result, but the second term in
Eq.~(\ref{check}) is not necessary in this check, again showing
its spurious nature.

As we have demonstrated, to construct light-cone sum rules
properly for the OPE given in Eq.~(\ref{tope}), {\it we must take only
the first term in Eq.~(\ref{check}), restrict the integral below
$S_0$, and perform the Borel transformation}.
%Alternatively, one
%may use directly $\rho^{\rm new} (s_1,s_2)$ without taking out the
%spurious contributions ``by hand''. Both methods are
%mathematically sensible.
Next, let us demonstrate how the OPE side looks
within this approach. Using the double Borel transformation
formula~\cite{Belyaev:1995zk}
\begin{eqnarray}
{\cal B}(M^2_1, -p_1^2) {\cal B}(M^2_2, -p_2^2) {(r-1)! \over
[s-v p^2_1 -(1-v) p_2^2]^r } &=& (M^2)^{2-r} e^{-s/M^2} \delta (v - v_0)\ ,
\nonumber \\
&& (r=1,2,3\cdot\cdot\cdot)\ 
\label{dformula}
\end{eqnarray}
we obtain, from the first term of Eq.(\ref{check}),
\begin{eqnarray}
{\cal B} (M_1^2,-p_1^2) {\cal B} (M_2^2, -p_2^2)
&&\int^1_0 dv~ v^k~  \int^{S_0}_0 ds_2 {1 \over (n-2)!}
{s_2^{n-2} \over s_2-A} \nonumber \\
%[\Pi^{\rm ope} - \Pi^{\rm cont}]
&=&{1\over \beta} {v_0^k \over (n-2)~!}
\left (-{\partial \over \partial \beta} \right )^{n-2}
\left (-{1\over \beta}~ e^{-s \beta} \right ) \Bigg |^{S_0}_0\ ,
\label{fc}
\end{eqnarray}
where we have defined
\begin{eqnarray}
\beta = {1 \over M^2} \;; \quad {1 \over M^2} =
{1\over M_1^2} + {1\over M_2^2}
\;; \quad  v_0 = {M_2^2 \over M_1^2 + M_2^2}\ .
\end{eqnarray}
The two Borel masses are often taken to be equal, i.e.
$M^2_1=M^2_2$~\cite{Belyaev:1995zk}, so that $v_0 = 1/2$. This implies 
that in the final form of a LCQSR, QCD inputs contain the wave
functions at the middle point $\varphi(1/2)$, i.e., the
probability for quark and antiquark to equally share the momentum
of the external field. By directly performing the derivative
$\left (-{\partial \over
\partial \beta} \right )^{n-2}$ in Eq.~(\ref{fc}), we obtain the
main result of this work,
\begin{eqnarray}
v_0^k~(M^{2})^n~E_{n-2}~({S_0 / M^2}).~~ (n\ge 2) \ 
\label{fcope}
\end{eqnarray}
Here we have defined the continuum factor as $E_n (x\equiv
S_0/M^2) = 1-(1+x+\cdot \cdot\cdot +x^n/n!)e^{-x}$. Note, the
factor $v_0^k (M^{2})^n$ is just twice the Borel transform of the
OPE given in Eq.~(\ref{tope}) without the continuum subtraction. We stress
that Eq.~(\ref{fcope}) is the correct expression to appear in the
final sum rule when the continuum is properly subtracted for the
OPE of Eq.~(\ref{tope}).

In contrast, in usual light-cone sum rules, the OPE after the
Borel transformation contains~\footnote{Most works on light-cone
sum rules do not show a clear derivation of this formula, but their
final OPE contains this continuum factor. [See for example
Eq.(5.13) in Ref.~\citen{Balitsky:ry}, the discussion on p.162 of
Ref.~\citen{Braun:1988qv}, or  Eq.(27) in
Ref.~\citen{Halperin:1992gt}. A technical derivation can also be
found in the appendix of Ref.~\citen{Aliev:2000rc}.]}
\begin{eqnarray}
v_0^k~(M^{2})^n~E_{n-1}~({S_0 / M^2}).~ (n\ge 2)\ 
\label{usualope}
\end{eqnarray}
This formula can be obtained from our approach by adding to our
result the term for $i=k$ with $s_2 =S_0$ in the second term 
involving the summation in Eq.~(\ref{check}). 
This additional (but spurious)
term becomes
\begin{eqnarray}
-\int^1_0 dv~v^k {1 \over (n-1)!} {S_0^{n-1} \over (S_0 -A)}\ .
\end{eqnarray}
Under the double Borel transformation given in Eq.~(\ref{dformula}),
this becomes
\begin{eqnarray}
-v^k_0 S_0^{n-1} M^2 e^{-S_0 /M^2}\ .
\end{eqnarray}
If this is added to our result Eq.~(\ref{fcope}), we exactly obtain
Eq.~(\ref{usualope}). This means that the usual LCQSR formula
Eq.~(\ref{usualope}) contains spurious continuum.
Furthermore, within the common method, it is not clear why
one can simply drop the continuum contribution from the terms with $i \ne k$
in Eq.~(\ref{check}).

As far as mathematical form is concerned,
Eq.~(\ref{usualope}) differs from our formula Eq.~(\ref{fcope})
only slightly. For a given power of Borel mass $(M^2)^n$,
Eq.~(\ref{usualope}) contains the continuum factor $E_{n-1}$ while our
formula Eq.~(\ref{fcope}) contains $E_{n-2}$. However, this slight
difference affects the final results substantially. We 
discuss this in the next section.

\section{Effects on existing sum rule analysis}
\label{sec:effect}

In this section, we discuss how the difference in the continuum
factor changes the predictions of previous LCQSR calculations. As
far as the $D^*D\pi$ coupling is concerned, our
prescription~\cite{Kim:2002xc} yields a value comparable to 
experiment~\cite{Anastassov:2001cw}, whereas
the previous light-cone sum rule calculation gives
a much smaller value~\cite{Belyaev:1995zk}. The changes
resulting from implementation of our prescription
will be most
effective in the sum rules that use a high-dimensional current,
for example, the sum rule for a nucleon magnetic
moment~\cite{Braun:1988qv} or a pion-nucleon
coupling~\cite{Zhu:1998am,Braun:1988qv}. Also, it will probably
affect the other baryon sum rules. In these sum rules, the
continuum threshold is $S_0 \sim 2 $ GeV$^2$, which corresponds
to the squared mass of the Roper resonance. The leading term in the OPE 
after the
Borel transformation contains the Borel mass $M^6$. For this term,
our result Eq.~(\ref{fcope}) suggests that the continuum factor $E_1
(x\equiv S_0/M^2) = 1 -(1+x)e^{-x}$ should be multiplied, while 
the conventional
light-cone sum rules~\cite{Zhu:1998am,Braun:1988qv} contain the
factor $E_2 = 1 -(1+x +x^2/2)e^{-x}$. At a typical Borel mass $M^2
\sim 1$ GeV$^2$, we have $E_1 \sim 0.6$, while $E_2 \sim 0.32$, only about
half of $E_1$. For an OPE leading to $M^4$, our continuum factor
is $E_0 \sim 0.86$, while the usual light-cone sum rules gives $E_1 \sim
0.6$, about 30 \% lower. Thus, Eq.~(\ref{usualope}) 
suppresses the perturbative contributions too strongly.
It is precisely this kind of suppression that leads to a 
value of the $D^*D\pi$ coupling that is much smaller than the experimental 
value.

\begin{figure}
\caption{ The nucleon magnetic moments ($F_2^p=\mu_p-1$ and
$F_2^n=\mu_n$). In (a), the solid curves 
qualitatively reproduce Fig.2. of Ref.~\citen{Braun:1988qv}.
The dashed curves represent the result when the continuum factors are corrected
in the manner described in the text. In (b), the leading term in
the OPE (dashed curve indicated by $f_1$) without the continuum,
with our continuum factor (dot-dashed curve indicated by $f_1 E_0$),
and with the usual light-cone continuum factor (dot-dashed curve 
indicated by $f_1 E_1$) are plotted. Also shown by the solid curve is the 
total OPE from Ref.~\citen{Braun:1988qv}}

\label{fig1}

\setlength{\textwidth}{6.1in}   % adjustable for camera-ready copy.
\setlength{\textheight}{9.in}  % for now.
%\begin{figure}
\centerline{%
\vbox to 2.4in{\vss
   \hbox to 3.3in{\includegraphics{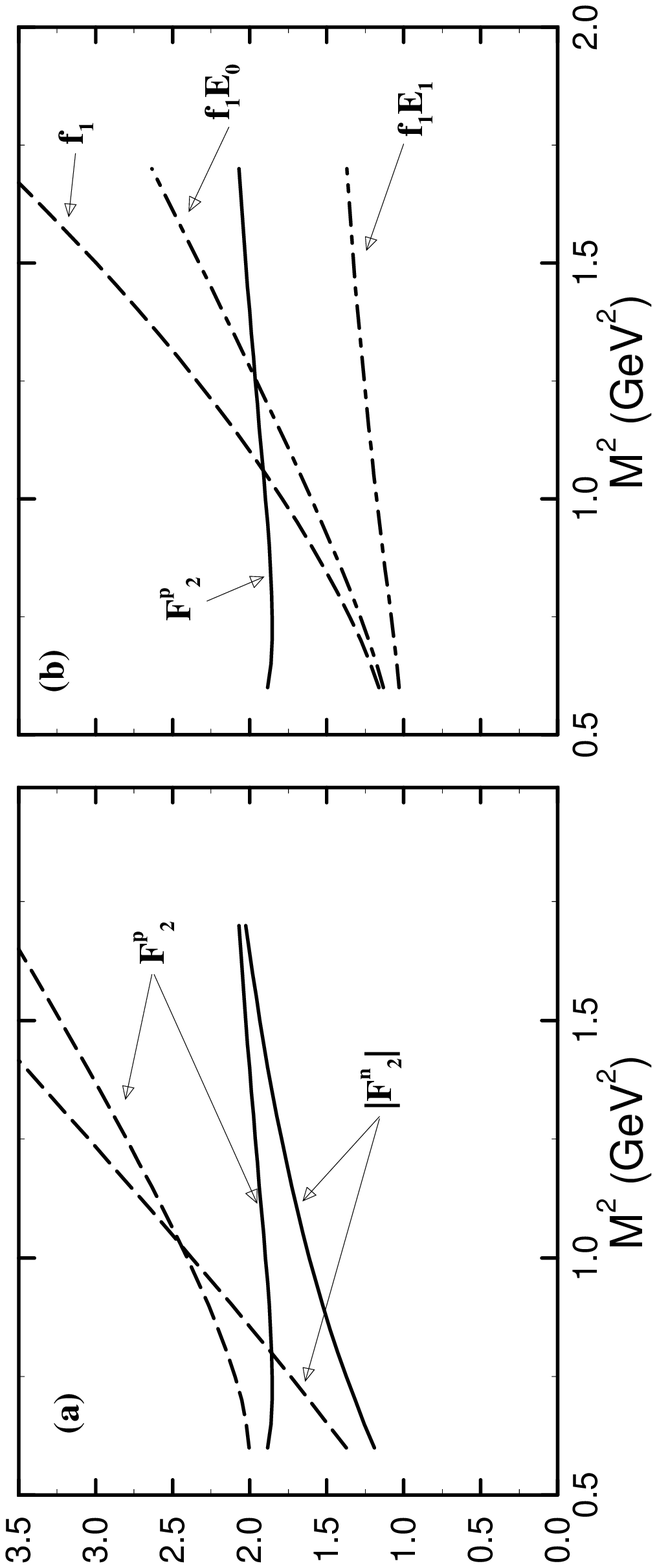}\hss}}
}
%\bigskip
%\vspace{20pt}
%\eject
\end{figure}

The first example to investigate the effect of the new prescription
is the calculation of the nucleon magnetic
moments~\cite{Braun:1988qv}. For this purpose, we simply take Eq.~(14) of
Ref.\citen{Braun:1988qv} and reproduce Fig.2 of that work
denoted by the solid curves in
Fig.~\ref{fig1} (a) here. When the sum rule is changed according to our
prescription, $E_1 \rightarrow E_0$ and $E_0 \rightarrow 1$, we
obtain the dashed curves in Fig.~\ref{fig1} (a). Depending on the
continuum factors, we clearly obtain quite different Borel
curves. It can be seen that the Borel stability of the solid curves
comes purely from the continuum factor, and, therefore, the
prediction is not stable with respect to variation of the continuum
threshold. To show this, we plot the leading term of the OPE {\it
without the continuum factor} in Fig.~\ref{fig1} (b) with the dashed curve 
(indicated by $f_1$). The magnitude of the rest of the OPE is approximately 
0.7 (not shown). As a reference curve, we again plot the
Borel curve (solid curve) for $F^p_2$. The dashed curve is already
above the total OPE (the solid curve indicated by $F_2^p$)
for $M^2 \ge 1.1$. The usual continuum factor containing the
spurious contribution lowers the curve substantially, as shown by
the dot-dashed curve [denoted as $f_1 E_1$ in Fig.~\ref{fig1} (b)],
which suppresses the perturbative part too strongly. The degree of
suppression depends on the Borel mass, but there is more than 50 \%
reduction for $M^2 \ge 1$ GeV$^2$, which results from only a simple
modeling of higher resonance contributions. Even if we restrict
the continuum contribution to be less than 50 \%, we still
cannot obtain a Borel window around 1 GeV$^2$, indicating that the
result is extremely  
sensitive to the continuum threshold. On the other
hand, as shown by the other dot-dashed curve (denoted by $f_1
E_0$), our prescription does not suppress the perturbative
part too strongly.

Another example to investigate is the pion-nucleon coupling 
calculation using the
nucleon two-point correlation function with a pion within the
light-cone sum rule approach~\cite{Zhu:1998am,Braun:1988qv},
\begin{eqnarray}
\Pi (q,p)=i \int d^4 x~ e^{i q \cdot x} \langle 0 | T[J_N (x)
{\bar J}_N (0)]| \pi (p) \rangle \ .
\label{cor}
\end{eqnarray}
In Ref.~\citen{Braun:1988qv}, light-cone sum rules are constructed
for the
$i\gamma_5 \fslash {p}$ and $i \gamma_5 \fslash{q} \fslash{p}$
Dirac structures~\footnote{ It should be noted that the $i\gamma_5
\fslash{q} \fslash{p}$ structure sum rule  is not independent of 
the $i\gamma_5$ Dirac structure. Since $i\gamma_5 \fslash{q}
\fslash{p}  = i\gamma_5 p\cdot q + \gamma_5 \sigma_{\mu \nu} q^\mu
p^\nu$, one actually needs to construct a sum rule for the
structure $\gamma_5 \sigma_{\mu \nu} q^\mu p^\nu$.} from the
correlation function Eq.~(\ref{cor}). They compared the two
sum rules and extracted the twist-2 pion wave function
at the symmetric point $\varphi_\pi (1/2)$, using
the experimental
pion-nucleon coupling $g_{\pi N} \sim 13.5$ as an input.
The
solid curves in Fig.~\ref{fig2} qualitatively reproduce the result
of Ref.~\citen{Braun:1988qv}. When the continuum factors are
changed according to our prescription given in Eq.~(\ref{fcope}), we obtain
the dashed curves, which are substantially lower than the solid
curves, again showing very large modifications of the sum rule results.

The OPE calculation for the $i\gamma_5 \fslash {p}$ sum
rule of Ref.~\citen{Braun:1988qv} was improved  by Zhu {\it
et al.}~\cite{Zhu:1998am}, who claimed that there are missing OPE
terms in Ref.~\citen{Braun:1988qv}. Even in Ref.~\citen{Zhu:1998am}, 
however, the spurious
continuum is very large. We simply take the formula Eq.~(23) of
Ref.~\citen{Zhu:1998am} and reproduce Fig.2 of
Ref.~\citen{Zhu:1998am} for $\varphi_\pi (1/2) \sim 1.5$ as shown
by the thick solid curve in Fig.~\ref{fig3} ($S_0 =2.25$ GeV$^2$).
However, if we simply take a slightly higher threshold, $S_0 = 2.75$
GeV$^2$, we obtain the thin solid curve. As expected, there is very 
strong
sensitivity to the continuum, and therefore the result of
Ref.~\citen{Zhu:1998am} is not conclusive. When the
continuum factors are corrected according to our formula
given in Eq.(\ref{fcope}), we obtain the thick dashed curve with which we
cannot conclude that their result $\varphi_\pi (1/2) \sim 1.5$ is
consistent with the pion-nucleon coupling $g_{\pi N} \sim 13.5$.

As we have shown in these examples, correcting continuum factors
substantially changes the results of previous light-cone sum rule
calculations.  
The failure to reproduce the known phenomenological
parameters may indicate that the approach adopted in
Refs.~\citen{Braun:1988qv} and \citen{Zhu:1998am} is not optimal for the
investigation of the baryon properties of interest. In particular,
the $i\gamma_5 \fslash {p}$ sum rule is known to be very sensitive
to the continuum
threshold~\cite{Kim:1998ir,Doi:2000dx}.  
Higher resonances with different parities add up to constitute
a very large continuum. Thus, it is not realistic to predict the coupling
after subtracting more than 50\% from the total strength
by a simple modeling of the continuum. 
Instead, one may need to
consider the $\gamma_5 \sigma_{\mu\nu}$ structure in the  
investigation of the $\pi NN$ coupling. In this case, higher
resonances with different parities cancel, and the resulting sum
rules are less sensitive to the continuum threshold. Similar
changes can be expected from other light-cone sum rules.
Therefore, it is important to re-analyze previous light-cone sum
rules using the continuum factor that we have considered in this
work.

In summary, a proper method of subtracting the continuum contribution
in light-cone QCD sum rules has been demonstrated in the work. In
particular, by closely looking into the double dispersion relation
and QCD duality, we have isolated the spurious contributions in
LCQSR. They are spurious because (1) they belong to subtraction
terms in the dispersion relation, and (2) in the limit of vanishing 
external momentum, they are precisely the terms that do not match
with that obtained using the single dispersion relation. We then proposed
the proper continuum factors to appear in the sum rules
for a given OPE.  We found that the continuum factor in
this approach is slightly different from the usual one appearing
in LCQSR, but the effect of this difference is  found to be enormous. 
It has been
demonstrated that the conclusions are altered greatly by 
this modification.

\section*{Acknowledgements}
The work of Hungchong Kim was supported by the
Korea Research Foundation Grant KRF-2002-015-CP0074.

%\appendix
%\section{First Appendix} %Empty argument \section{} yields `Appendix'. 
%
%\section{Second Appendix}

\eject

\begin{figure}
\caption{ The twist-2 pion wave function at the middle point
$\varphi_\pi (1/2)$ corresponding to the result of Ref.~\citen{Braun:1988qv}. 
Again, the
solid curves reproduce Fig.4 of Ref.~\citen{Braun:1988qv}, and
the dashed curves are those obtained when the continuum is corrected. }
\label{fig2}
\end{figure}
%\eject

\setlength{\textwidth}{6.1in}   % adjustable for camera-ready copy.
\setlength{\textheight}{9.in}  % for now.
\begin{figure}
\centerline{%
\vbox to 2.4in{\vss
   \hbox to 3.3in{\includegraphics{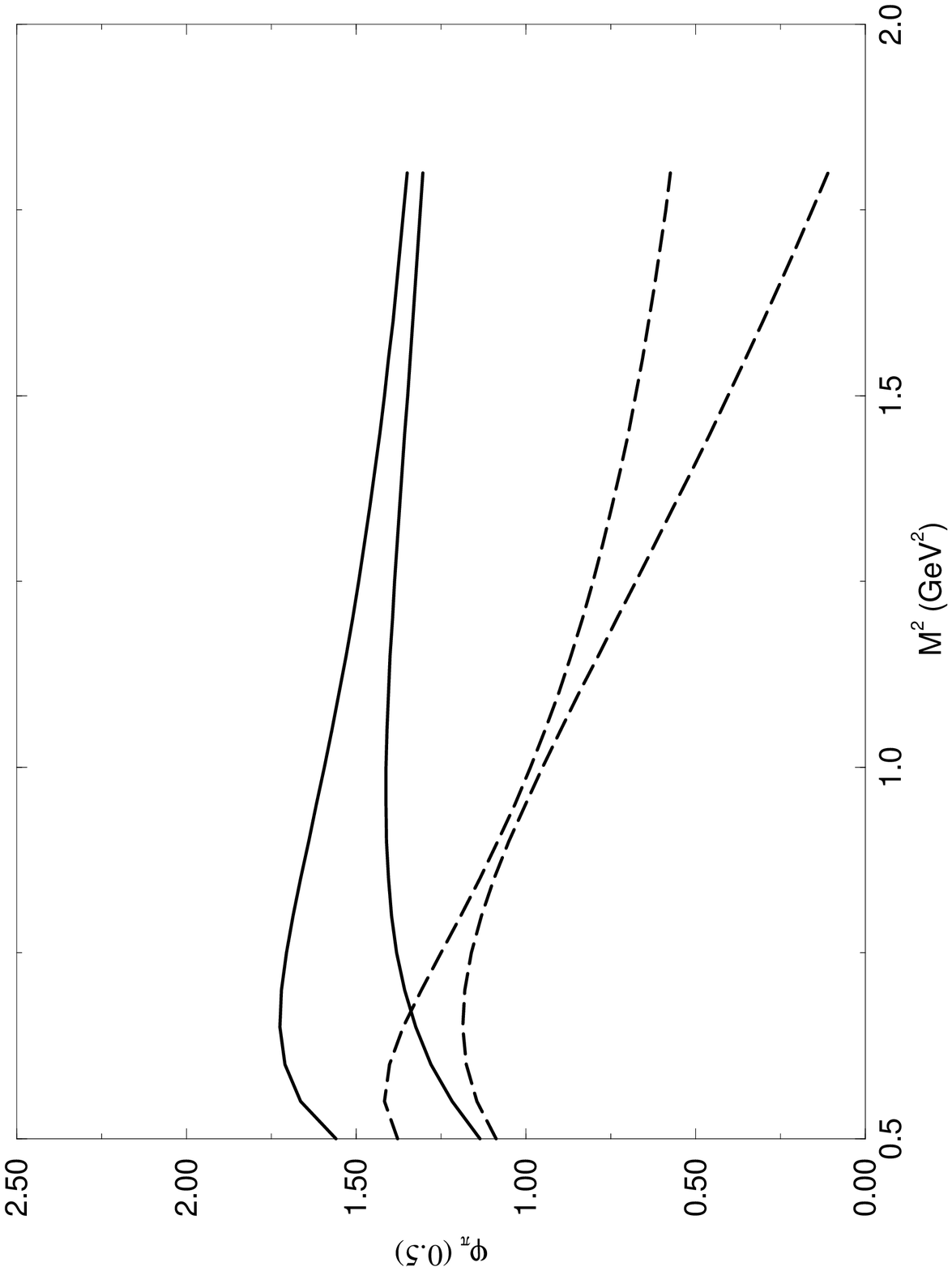}\hss}}
}
%\bigskip
%\vspace{400pt}
\end{figure}

\eject

\begin{figure}
\caption{ The Borel curves of the pion-nucleon coupling corresponding to
the result in  Ref.~\citen{Zhu:1998am}. The thick solid curve reproduces the
result of Ref.~\citen{Zhu:1998am} while the thick dashed curve is
obtained when the continuum factor is corrected according to our
prescription. The corresponding thin curves are obtained when the
continuum threshold is shifted by only $0.5$ GeV$^2$, showing that the
result is very sensitive to the continuum threshold.}

\label{fig3}

\setlength{\textwidth}{6.1in}   % adjustable for camera-ready copy.
\setlength{\textheight}{9.in}  % for now.
%\begin{figure}
\centerline{%
\vbox to 2.4in{\vss
   \hbox to 3.3in{\includegraphics{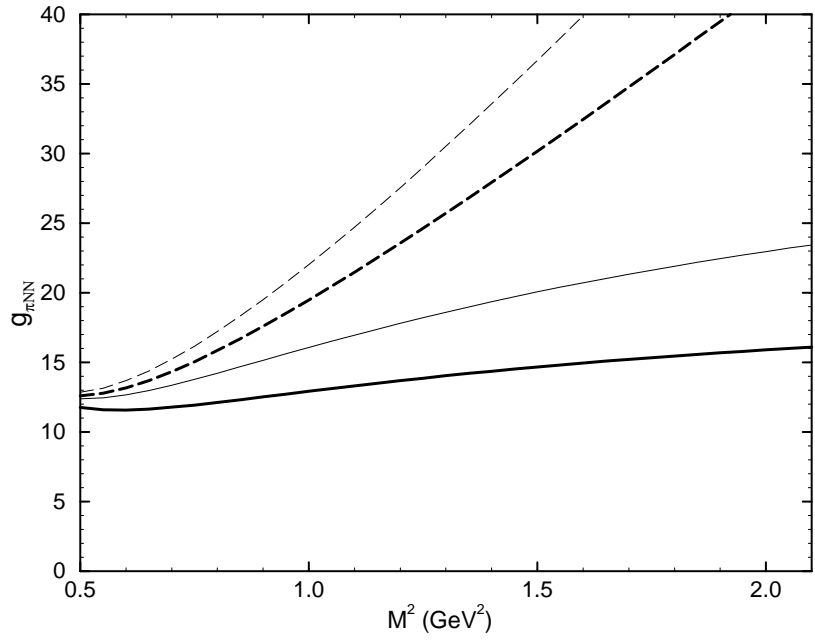}\hss}}
}
%\bigskip
%\eject
\end{figure}


\begin{thebibliography}{99}
%%%%%%%%%%%%%%%%%%%%%%%%%%%%%%%%%%%%%%%%%%%%%%%%%%%%%%%%%%%%%
% Some macros are available for the bibliography:
%  o for general use
%    \JL : general journals                 \andvol : Vol (Year) Page
%  o for individual journal 
%    \AJ   : Astrophys. J.           \NC         : Nuovo Cim.
%    \ANN  : Ann. of Phys.           \NPA, \NPB  : Nucl. Phys. [A,B]
%    \CMP  : Commun. Math. Phys.     \PLA, \PLB  : Phys. Lett. [A,B]
%    \IJMP : Int. J. Mod. Phys.      \PRA - \PRE : Phys. Rev. [A-E]     
%    \JHEP : J. High Energy Phys.    \PRL        : Phys. Rev. Lett.
%    \JMP  : J. Math. Phys.          \PRP        : Phys. Rep.
%    \JP   : J. of Phys.             \PTP        : Prog. Theor. Phys.     
%    \JPSJ : J. Phys. Soc. Jpn.      \PTPS       : Prog. Theor. Phys. Suppl.
% Usage:
%  \PRD{45,1990,345}          ==> Phys.~Rev.\ \textbf{D45} (1990), 345
%  \JL{Nature,418,2002,123}   ==> Nature \textbf{418} (2002), 123
%  \andvol{B123,1995,1020}    ==> \textbf{B123} (1995), 1020
%%%%%%%%%%%%%%%%%%%%%%%%%%%%%%%%%%%%%%%%%%%%%%%%%%%%%%%%%%%%%
  


%\cite{Shifman:bx}
\bibitem{Shifman:bx}
M.~A.~Shifman, A.~I.~Vainshtein and V.~I.~Zakharov,
%``QCD And Resonance Physics. Sum Rules,''
Nucl.\ Phys.\ B {\bf 147} (1979) 385.
%%CITATION = NUPHA,B147,385;%%


%\cite{Shifman:by}
\bibitem{Shifman:by}
M.~A.~Shifman, A.~I.~Vainshtein and V.~I.~Zakharov,
%``QCD And Resonance Physics: Applications,''
Nucl.\ Phys.\ B {\bf 147} (1979) 448.
%%CITATION = NUPHA,B147,448;%%


%\cite{Reinders:1984sr}
\bibitem{Reinders:1984sr}
L.~J.~Reinders, H.~Rubinstein and S.~Yazaki,
%``Hadron Properties From QCD Sum Rules,''
Phys.\ Rept.\  {\bf 127} (1985) 1. 
%%CITATION = PRPLC,127,1;%%


%\cite{Kim:1998wp}
\bibitem{Kim:1998wp}
H.~Kim, S.~H.~Lee and M.~Oka,
%``QCD sum rule calculation of the pi N N coupling - revisited,''
Phys.\ Lett.\ B {\bf 453} (1999) 199, nucl-th/9809004.
%%CITATION = NUCL-TH 9809004;%%

%\cite{Kim:1999ue}
\bibitem{Kim:1999ue}
H.~Kim,
%``pi N N coupling determined beyond the chiral limit,''
Eur.\ Phys.\ J.\ A {\bf 7} (2000) 121, nucl-th/9904049.
%%CITATION = NUCL-TH 9904049;%%

%\cite{Kim:1998ir}
\bibitem{Kim:1998ir}
H.~Kim, S.~H.~Lee and M.~Oka,
%``Two-point correlation function with pion in QCD sum rules,''
Phys.\ Rev.\ D {\bf 60} (1999) 034007, nucl-th/9811096.
%%CITATION = NUCL-TH 9811096;%%

%\cite{Kim:1999tc}
\bibitem{Kim:1999tc}
H.~Kim, T.~Doi, M.~Oka and S.~H.~Lee,
%``Meson baryon couplings and the F/D ratio from QCD sum rules,''
Nucl.\ Phys.\ A {\bf 662} (2000) 371, nucl-th/9909007.
%%CITATION = NUCL-TH 9909007;%%

%\cite{Kim:2000ty}
\bibitem{Kim:2000ty}
H.~Kim, T.~Doi, M.~Oka and S.~H.~Lee,
%``The F/D ratio and meson baryon couplings from QCD sum rules. II,''
Nucl.\ Phys.\ A {\bf 678} (2000) 295, nucl-th/0002011.
%%CITATION = NUCL-TH 0002011;%%

%\cite{Doi:2000dx}
\bibitem{Doi:2000dx}
T.~Doi, H.~Kim and M.~Oka,
%``Pertinent Dirac structure for QCD sum rules of meson baryon coupling  constants,''
Phys.\ Rev.\ C {\bf 62} (2000) 055202, nucl-th/0004065.
%%CITATION = NUCL-TH 0004065;%%

%\cite{Aliev:1999ce}
\bibitem{Aliev:1999ce}
T.~M.~Aliev and M.~Savci,
%``Pion baryon coupling constants in light cone {QCD} sum rules,''
Phys.\ Rev.\ D {\bf 61} (2000) 016008, hep-ph/9904296.
%%CITATION = HEP-PH 9904296;%%

%\cite{Zhu:1998am}
\bibitem{Zhu:1998am}
S.~L.~Zhu, W.~Y.~Hwang and Y.~B.~Dai,
%``The pi N N and pi N N(1535) couplings in QCD,''
Phys.\ Rev.\ C {\bf 59} (1999) 442, nucl-th/9809033.
%%CITATION = NUCL-TH 9809033;%%




%\cite{Belyaev:1995zk}
\bibitem{Belyaev:1995zk}
V.~M.~Belyaev, V.~M.~Braun, A.~Khodjamirian and R.~Ruckl,
%``D* D pi and B* B pi couplings in QCD,''
Phys.\ Rev.\ D {\bf 51} (1995) 6177, hep-ph/9410280.
%%CITATION = HEP-PH 9410280;%%

%\cite{Ioffe:ju}
\bibitem{Ioffe:ju}
B.~L.~Ioffe and A.~V.~Smilga,
%``Nucleon Magnetic Moments And Magnetic Properties Of Vacuum In QCD,''
Nucl.\ Phys.\ B {\bf 232} (1984) 109 .
%%CITATION = NUPHA,B232,109;%%

%\cite{Aliev:2000rc}
\bibitem{Aliev:2000rc}
T.~M.~Aliev, A.~Ozpineci and M.~Savci,
%``Magnetic moments of Delta baryons in light cone QCD sum rules,''
Nucl.\ Phys.\ A {\bf 678}(2000) 443, hep-ph/0002228.
%%CITATION = HEP-PH 0002228;%%

%\cite{Balitsky:ry}
\bibitem{Balitsky:ry}
I.~I.~Balitsky, V.~M.~Braun and A.~V.~Kolesnichenko,
%``Radiative Decay Sigma+ $\to$ P Gamma In Quantum Chromodynamics,''
Nucl.\ Phys.\ B {\bf 312} (1989) 509.
%%CITATION = NUPHA,B312,509;%%

%\cite{Braun:1988qv}
\bibitem{Braun:1988qv}
V.~M.~Braun and I.~E.~Halperin,
%``QCD Sum Rules In Exclusive Kinematics And Pion Wave Function,''
Z.\ Phys.\ C {\bf 44} (1989) 157.  
%%CITATION = ZEPYA,C44,157;%%


%\cite{Braun:1989iv}
\bibitem{Braun:1989iv}
V.~M.~Braun and I.~E.~Halperin,
%``Conformal Invariance And Pion Wave Functions Of Nonleading Twist,''
Z.\ Phys.\ C {\bf 48} (1990) 239.  
%%CITATION = ZEPYA,C48,239;%%

%\cite{Halperin:1992gt}
\bibitem{Halperin:1992gt}
I.~E.~Halperin,
%``Vertex QCD sum rules, quark model and pion wave function,''
Z.\ Phys.\ C {\bf 56} (1992) 615.
%%CITATION = ZEPYA,C56,615;%%

%\cite{Kim:1999rg}
\bibitem{Kim:1999rg}
H.~Kim,
%``Comment on 'Determination of pion baryon coupling constants from QCD  sum rules',''
Phys.\ Rev.\ C {\bf 61} (2000) 019801, nucl-th/9903040.
%%CITATION = NUCL-TH 9903040;%%

%\cite{Ioffe:1995jt}
\bibitem{Ioffe:1995jt}
B.~L.~Ioffe,
%``On the three point vertex of hadron interaction with 
%external fields in QCD sum rules,''
Phys.\ Atom.\ Nucl.\  {\bf 58} (1995) 1408, hep-ph/9501319.
%%CITATION = HEP-PH 9501319;%%

%\cite{Krippa:1999de}
\bibitem{Krippa:1999de}
B.~Krippa and M.~C.~Birse,
Phys.\ Rev.\ C {\bf 61} (2000) 019802, nucl-th/9904003.
%%CITATION = NUCL-TH 9904003;%%


%\cite{Kim:1999ur}
\bibitem{Kim:1999ur}
H.~Kim,
%``Use of double dispersion relation in QCD sum rules with external  fields,''
Prog.\ Theor.\ Phys.\  {\bf 103} (2000) 1001, nucl-th/9906081.
%%CITATION = NUCL-TH 9906081;%%



%\cite{Kim:2002xc}
\bibitem{Kim:2002xc}
H.~Kim,
%``Re-analysis of the D* D pi coupling in the light-cone QCD sum rules,''
hep-ph/0206170.
{\it To be published in Journal of the Korean Physical Society}
%%CITATION = HEP-PH 0206170;%%

\bibitem{Anastassov:2001cw}
A.~Anastassov {\it et al.}  [CLEO Collaboration],
%``First measurement of Gamma(D*+) and precision measurement of m(D*+) -  m(D0),''
Phys.\ Rev.\ D {\bf 65} (2002) 032003, hep-ex/0108043.

%\cite{Becirevic:1999fr}
\bibitem{Becirevic:1999fr}
D.~Becirevic and A.~L.~Yaouanc,
%``g-hat coupling (g(B* B pi), g(D* D pi)): A quark model with Dirac  equation,''
J.High Energy Phys. {\bf 9903} (1999) 021, hep-ph/9901431.
%%CITATION = HEP-PH 9901431;%%

%\cite{Abada:2002xe}
\bibitem{Abada:2002xe}
A.~Abada {\it et al.},
%``First lattice QCD estimate of the g(D* D pi) coupling,''
hep-ph/0206237.
%%CITATION = HEP-PH 0206237;%%



%\cite{Novikov:gd}
\bibitem{Novikov:gd}
V.~A.~Novikov, M.~A.~Shifman, A.~I.~Vainshtein and V.~I.~Zakharov,
%``Calculations In External Fields In Quantum Chromodynamics:. Technical Review (Abstract Operator Method, Fock-Schwinger Gauge),''
Fortsch.\ Phys.\  {\bf 32} (1985) 585.
%%CITATION = FPYKA,32,585;%%

%\cite{Beilin:1984pf}
\bibitem{Beilin:1984pf}
V.~A.~Beilin and A.~V.~Radyushkin,
%``Quantum Chromodynamic Sum Rules And J / Psi $\to$ Eta(C) Gamma Decay,''
Nucl.\ Phys.\ B {\bf 260} (1985) 61.
%%CITATION = NUPHA,B260,61;%%

%\cite{Ball:1993xv}
\bibitem{Ball:1993xv}
P.~Ball and V.~M.~Braun,
%``Next-to-leading order corrections to meson masses in the heavy quark effective theory,''
Phys.\ Rev.\ D {\bf 49} (1994) 2472, hep-ph/9307291.
%%CITATION = HEP-PH 9307291;%%

%\cite{Nesterenko:1982gc}
\bibitem{Nesterenko:1982gc}
V.~A.~Nesterenko and A.~V.~Radyushkin,
%``Sum Rules And Pion Form-Factor In QCD,''
Phys.\ Lett.\ B {\bf 115} (1982) 410.
%%CITATION = PHLTA,B115,410;%%

\end{thebibliography}
\end{document}